\documentclass[epj]{svjour}
\usepackage{graphics}
\usepackage{epsfig}
\usepackage{amssymb}
\usepackage{amsmath}
\usepackage{amsfonts}
\newcommand{\ar}{\arrowvert}

\newcommand{\da}{\dagger}

\newcommand{\be}{\begin{equation}}
\newcommand{\ee}{\end{equation}}
\newcommand{\ba}{\begin{eqnarray}}
\newcommand{\ea}{\end{eqnarray}}

\begin{document}
\def\Id{{\rm 1\kern-.3em I}}
\def\Psibar{\overset{\rule{2mm}{.2mm}}{\Psi}}
\def\ubar{\overset{\rule{2mm}{.2mm}}{u}}
\def\vbar{\overset{\rule{2mm}{.2mm}}{v}}
\def\ud{\mathrm{d}}
\def\qsl{q\hspace{-0.2cm}/}
\def\ksl{k\hspace{-0.2cm}/}
\def\psl{p\hspace{-0.2cm}/}
\def\mL{\mathcal{L}}
\def\mM{\mathcal{M}}
\def\mR{\mathcal{R}}

\title{Fermion family recurrences in the Dyson-Schwinger formalism}

\author{Felipe J. Llanes-Estrada\thanks{\emph{Contact email
\tt{fllanes@fis.ucm.es} } }, Tim Van Cauteren $ ^\da$ and \'Angel
P\'aramo Mart\'{\i}n
}
\institute{Depto. F\'{\i}sica Te\'orica I, Universidad Complutense de 
Madrid, Avda. Complutense s/n, 28040 Madrid, Spain.\\
$ ^\da$ (on leave from Ghent University, Belgium)}
\authorrunning{Llanes-Estrada, Van Cauteren and Paramo}
\titlerunning{Fermion family recurrences in DSE's}
\date{Received: date / Revised version: date}
%
\abstract{
We study the multiple solutions of the truncated propagator 
Dyson-Schwinger equation for a simple fermion theory with Yukawa 
coupling to a scalar field. Upon increasing the coupling constant $g$, 
other parameters being fixed, more than one non-perturbative solution
breaking chiral symmetry becomes possible and we find these numerically. 
These ``recurrences'' appear as a mechanism to generate 
different fermion generations as quanta of the same fundamental field in 
an interacting field theory, without assuming any composite structure. 
The number of recurrences or flavors is reduced to a question 
about the value of the Yukawa coupling, and has no special profound 
significance in the Standard Model.
The resulting mass function can have  one or more nodes
and the measurement that potentially detects them can be thought of  as 
a collider-based  
test of the virtual dispersion relation 
$E=\sqrt{p^2+M(p^2)^2}$ for the charged lepton member of each family. 
This 
requires three independent measurements of the charged lepton's  energy, 
three-momentum and off-shellness. We illustrate how this can be achieved 
for the (more difficult) case of the tau lepton.
} 
\maketitle
\section{Introduction}
\label{intro}

Why Nature has laid down exactly three fermion families in
the same representation of the Standard Model's gauge groups with
universal couplings to the gauge bosons remains to be explained. 
Many ideas have been presented in the literature and we quote some to 
exemplify the far-reaching implications of any experimental progress 
in the first question of what has been called the ``Fermion problem''.

``Democratic approaches'' \cite{Fritzsch:1994yj} are in general based on
the idea of equal fermion-to-Higgs Yukawa couplings. One then needs
mechanisms to generate specific lepton or quark mass patterns, see for 
example \cite{Dermisek:2003rw}.  Let us also recall the classic work of 
Froggat and Nielsen 
\cite{Froggatt:1978nt} that introduces the concept of ''horizontal 
flavor'' symmetry (nowadays often called Interfamily symmetry) by which 
the three families are degenerate or quasi-degenerate at some high scale 
beyond present experimental reach, due to unknown symmetry. At lower 
energies dynamical (renormalization group) effects amplify any small 
symmetry breaking term causing the ratios between masses that we see in 
current accelerators.
  In essence, the idea is that each of the elementary particle types, 
for example the charged leptons $(e,\mu,\tau)$ provide a representation 
of the posited interfamily symmetry. Since the symmetry is broken at the 
TeV scale and below, revealing it requires theoretical extrapolation 
from current experimental data. The same authors have followed-up with 
ideas based on  Anti-grand unification \cite{Froggatt:1996np}.

An alternative line of research assumes that the known elementary 
particles are really composite objects of more fundamental ``preons''
\cite{Pati:1984jf}. To date no experiment has revealed composite 
structure beyond the usual layers of quantum pairs involved in radiative 
corrections around seemingly point-like sources. 
One can argue that a new strong interaction binds preons and this might 
conceivably be revealed in future experimental efforts.

Further, dynamical breaking of chiral symmetry as a genuine quantum 
effect is well known in strong interaction physics  
\cite{Roberts:2003uk,Maris:2003vk}
and has also been invoked in flavor physics as an alternative to the 
tree-level Higgs mechanism, for example by Brauner and Hosek 
\cite{Brauner:2005hw}. 

However no one seems to have paid attention to 
the fact that the Dyson-Schwinger equations do provide one 
with several solutions depending on the strength of the coupling. In 
this work we call attention to this point and its potential relevance 
for the fermion problem. 
Each of the three fermions corresponds in this hypothesis to a quantum 
over a different vacuum (but in the covariant formalism one pursues the 
study of correlation functions, here propagators, and sidesteps the 
issue of the vacuum wave functional).
Far from attempting a 
complete theory, we will show the general features of 
the mechanism within a simple model of a fermion field coupled to a 
scalar boson through a Yukawa coupling. 

From this prospective, different flavored fermions, one 
corresponding to each family, are quanta of the same field, 
so the lagrangian can be more economically written. 
Each fermion is an elementary one-particle excitation on top of a vacuum 
that is a local minimum of the Hamiltonian. The 
excitations over the ground-state vacuum provide the lightest fermion 
family. Finally, at fixed coupling $g$, the Dyson-Schwinger equation
has a finite number of solutions, so a finite number of families arise.
The question on why three families becomes only a question of the value 
of $g$, the interaction coupling, and may have no special significance.

The equivalent of the propagator Dyson-Schwinger equation in a 
non-covariant framework is the well studied mass gap equation of 
potential models of QCD. The equation has been numerically solved for 
the harmonic oscillator potential \cite{Bicudo:1989sh}, 
\cite{LeYaouanc:1984dr}, the linear potential \cite{Adler:1984ri}, and 
the linear plus Coulomb potential 
\cite{Szczepaniak:1996gb,Llanes-Estrada:2001kr}. 
Bicudo, Ribeiro and Nefediev have systematically studied the excited 
solutions of this equation \cite{Bicudo:2002eu}. 
Our results seem compatible with this prior work.

\section{Dyson-Schwinger equation and its numerical solution}

We examine a simple Yukawa theory for one fermion field coupled to a 
real scalar field with Lagrangian density in Euclidean space
\be
\mathcal{L} =  \bar{\Psi} (i \not p-m_\Psi) \Psi + g \bar{\Psi} \Psi 
\phi
+ \frac{1}{2}\phi (\partial_\mu \partial^\mu -m_\phi^2) \phi  \ .
\ee

\begin{figure*}
\resizebox{0.75\textwidth}{!}{%
  \includegraphics{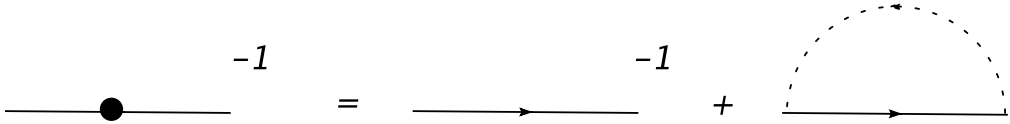}
}
\caption{Dyson-Schwinger equation for the fermion propagator in Yukawa 
theory, in the rainbow approximation.}
\label{DSE}
\end{figure*}

The Dyson-Schwinger equation for the fermion propagator in this theory 
is represented in figure \ref{DSE} in the rainbow approximation 
(neglecting the vertex dressing). 

To examine generic features of dynamical chiral symmetry breaking we can 
further ignore the running of the wavefunction renormalization and set 
$Z(p^2)=1$, a constant. This leaves one scalar equation for the 
fermion mass function that  reads
\be
M(p^2)= m_\Psi Z_2 + \int_0^\infty dq F(q,p) \frac{M(q^2)}{q^2+M^2(q^2)} 
\ .
\ee
In the chiral limit $m_\Psi \to 0$ this equation becomes homogeneous and 
accepts a chiral-symmetry preserving solution $M=0$ that continuously 
deforms into a soft-running mass form when $m_\Psi$ is not zero.
But for strong enough kernels $F$ it also admits other solutions that 
break chiral symmetry.

The kernel in this simple Yukawa theory is
\be \label{kernel}
F(q,p)= \frac{g^2 q^3}{4\pi^3} \int_{-1}^1 \sqrt{1-y^2} dy 
\frac{1}{(q-p)^2+m_\phi^2}
\ee
with $y$ the cosine of the polar angle in Euclidean four-dimensional 
space from $p$ to $q$.

To obtain excited solutions for $M$ as function of $p^2$ we apply an 
iterative linear method with different initial guesses with or without 
nodes for $M(p^2)$.
This method proceeds by examining linear deviations of the exact 
solution 
\be
M(p^2)=M_0(p^2)+\epsilon (p^2)
\ee
that yield
\ba
\epsilon (p^2)-\int_0^\infty dq F(q,p) \epsilon (q^2) \left(
\frac{q^2-M^2(q^2)}{(q^2+M^2(q^2))^2} \right) \\ \nonumber
=m_\Psi Z_2-M_0(p^2)+\int_0^\infty dq F(q,p) 
\frac{M_0(q^2)}{q^2+M_0^2(q^2)} \ 
.
\ea
This can in turn be discretized (simply discretizing $p$) and cast as a 
linear system
\be
A_{ij} \epsilon_j = (RHS)_i
\ee
and solved with standard numerical linear algebra tools. The kernels 
require only the calculation of 2-dimensional integrals that are also 
standard fare in modern computers. 

By naive dimensional analysis one sees that the integral needs UV 
regularization. We accomplish this by a simple cut-off method. Then the 
parameters should be chosen to run with the cutoff $g(\Lambda)$,
 $m_\psi (\Lambda)$ 
to ensure that the mass functions turn largely independent of $\Lambda$.  
In our results we display the dependence of the solution with $g$ at 
fixed $\Lambda$. The boson mass $m_\phi$ 
is chosen to be $1$ and sets the unit of the theory. 

\section{Numerical results and discussion}
\begin{figure*}
\psfig{figure=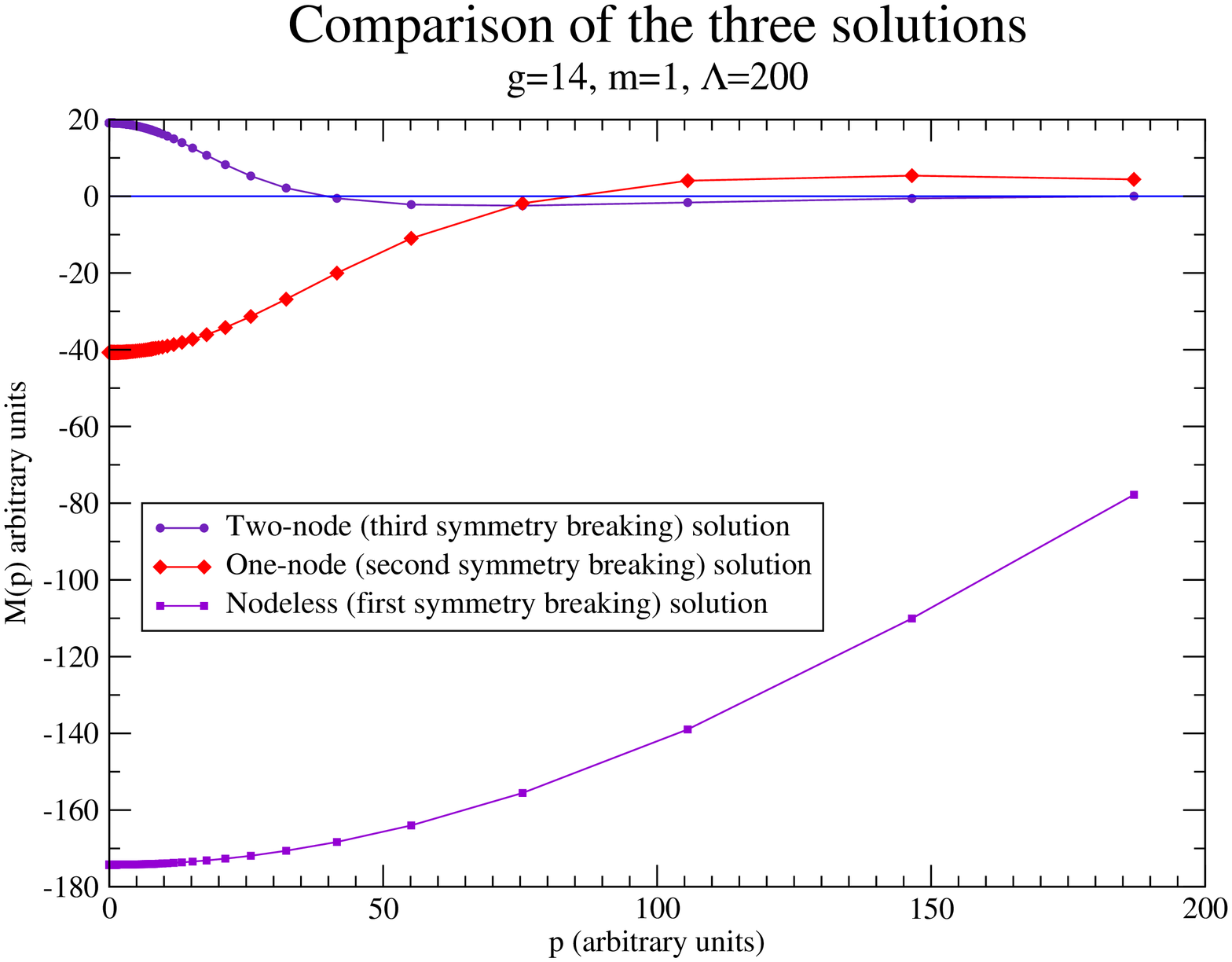,height=7in,angle=-90}
\caption{\label{tressoluciones}
We plot together the three solutions to the Dyson-Schwinger equation
for a fermion-scalar Yukawa coupling $g=14$, cutoff $\Lambda=200$, and
the scale set by the scalar mass being fixed to $m_\phi=1$. Note the 
equation for $M$ in the chiral limit (zero current fermion mass) cannot 
distinguish $M$ from $-M$ so the global sign of any one solution in 
this graph is irrelevant and for display purposes only. In particular 
all M(0) can be taken as positive.} 
\end{figure*}

\begin{figure*}
\psfig{figure=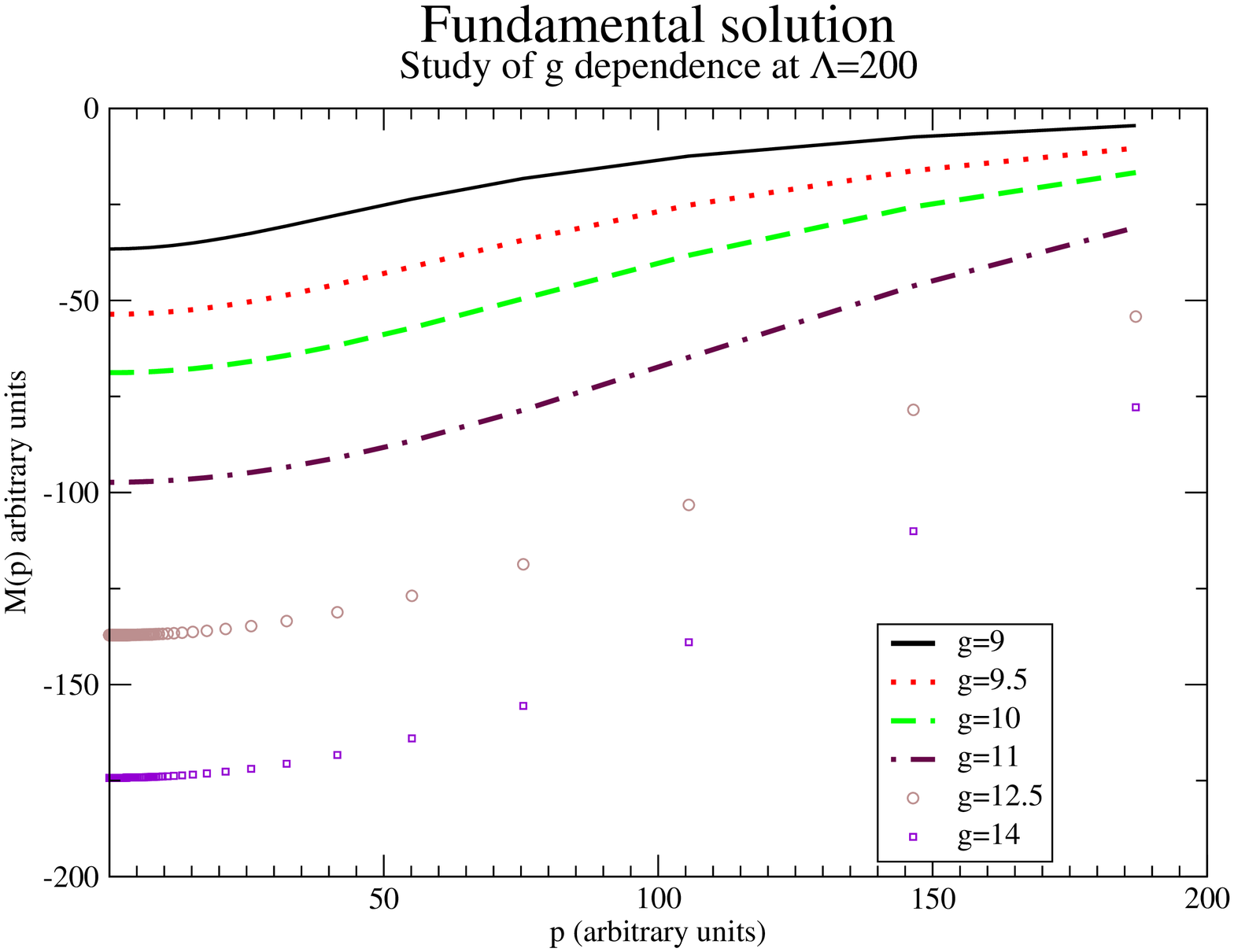,height=7in,angle=-90}
\caption{\label{groundsolution}
Nodeless solution for increasing values of the coupling 
constant ($m_\phi=1$ and $\Lambda=200$ are held fixed). 
Note how $M(0)$ grows steadily with $g$ (the sign of $M$ is
irrelevant).
}
\end{figure*}

In this section we present the chiral-symmetry breaking conventional and 
new excited state solutions.
By running our computer program incrementing $g$ we meet critical $g$ 
values $g_0$, $g_1$, ...$g_n$ above which there are exactly $n+1$ 
solutions, the last with $n$ nodes, that we find numerically. For 
example the first solution appears for a critical $g_0\simeq 6.8$ (of 
course this number runs with the cutoff). 

We show in figure \ref{tressoluciones} the three solutions obtained with
$g=14$, $\Lambda=200$. 

Then in figure \ref{groundsolution} we plot the nodeless solution at 
fixed cutoff incrementing $g$ sequentially to show the dependence of the 
mass function on the coupling. 

Only the nodeless solution has been widely used in past literature for 
its applications in hadron physics. 
The others are sometimes rejected \cite{kizillersu} on the basis of 
``wrong'' UV asymptotic behavior, arguing that they do not match with 
conventional mass running in perturbation theory, and artificially 
imposing a boundary condition that discards all but one solution. 

We point out here that this mismatch poses no problem. One should 
interpret these solutions as the one-particle excitations above 
different extrema of the Hamiltonian or vacuum replicae. Therefore they 
cannot all be matched to the same perturbation theory around one 
particular vacuum.

If one insists on viewing them from the one vacuum that connects 
smoothly to perturbation theory, then they have to be written down as 
complicated collective states in terms of the one-particle solutions 
over this vacuum.

In the particular case of the Yukawa theory, the angular integral in eq. 
\ref{kernel} can be performed analytically \cite{Llanes-Estrada:2003ha}, 
but we prefer keeping a two-dimensional integral thinking of future 
work with more complicated model Lagrangians.

In agreement with the findings of \cite{Bicudo:2002eu}, we see that the 
higher excited vacua in this covariant Euclidean formulation present 
zeroes in the mass function. A version of the Sturm-Liouville theorem 
must be at work (this is not surprising since the Hamiltonian is 
hermitian).

\begin{figure*}
\psfig{figure=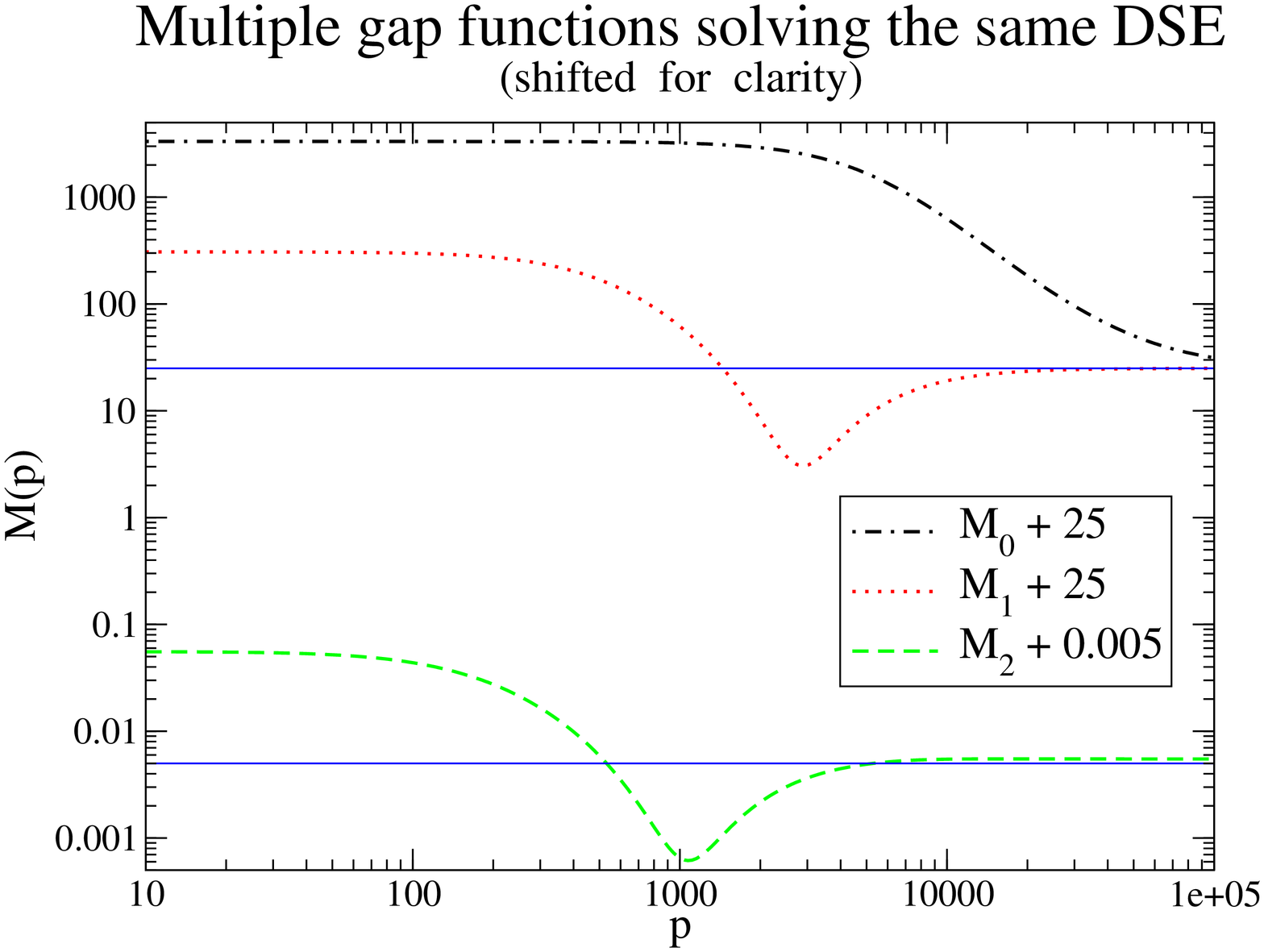,height=7in,angle=-90}
\caption{\label{newsolplot}
Three solutions to the gap equation in Yukawa theory obtained with
$m_\phi=100$, $\Lambda=10^4$, $m_\psi=5\cdot 10^{-4}$, $g=50$,
confirming that the nodeless solution is higher in mass, and that a
sizeable hierarchy of masses is possible. To represent them in a
log-log plot, $M_1$ and $M_2$ being negative in part of their domain,
the solutions have been shifted by a constant amount to positive
values.}
\end{figure*}

\begin{figure*}
\psfig{figure=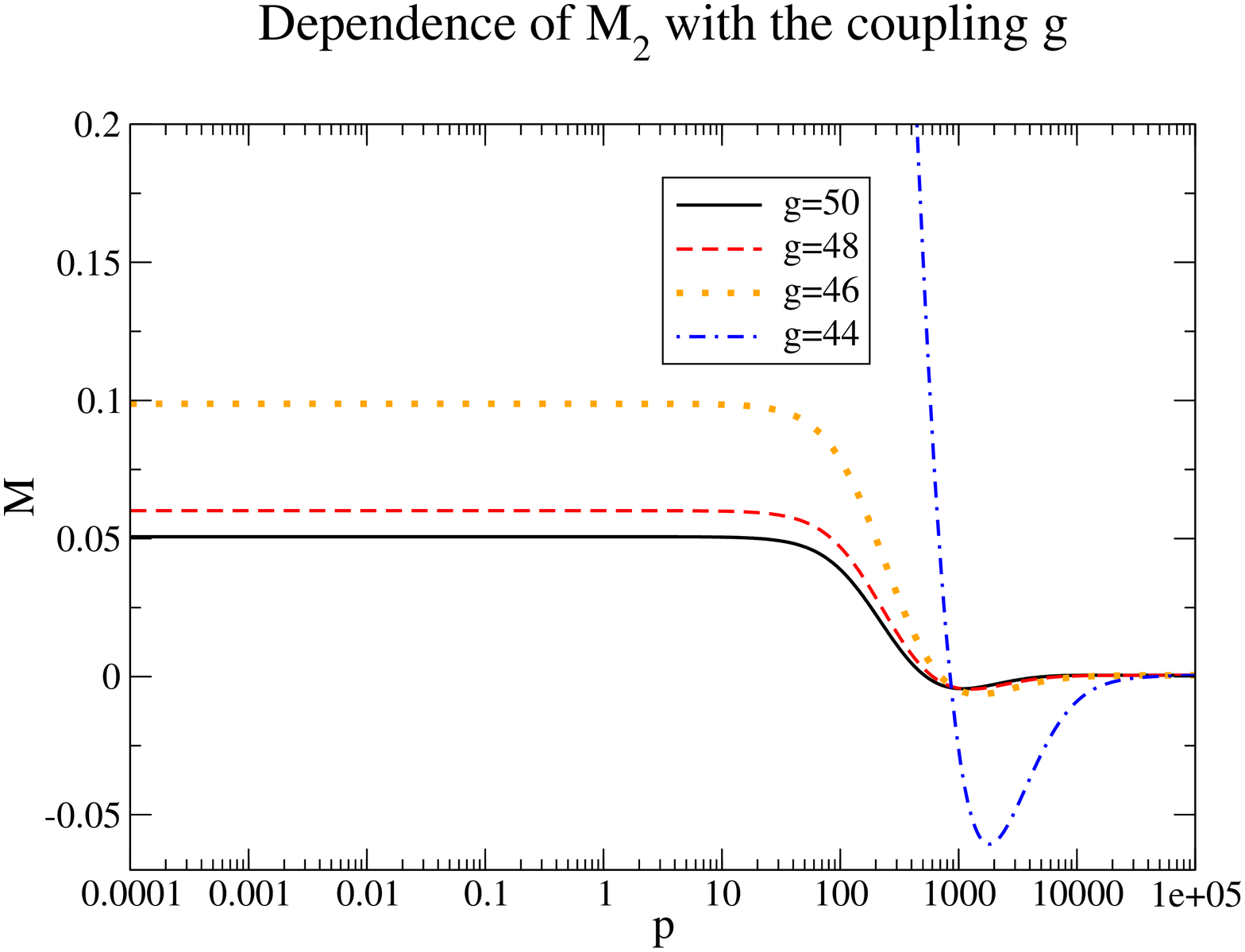,height=7in,angle=-90}
\caption{\label{M2runswithg} Dependence of $M_2$ with the coupling 
constant at fixed $m_\phi=100$, $\Lambda=10^4$, $m_\psi=5\cdot 10^{-4}$.
Upon comparing with figure \ref{groundsolution} we see that the 
qualitative behavior when the solution approaches the critical coupling 
and is about to cease existing, here $g=43$, is quite different for the 
ground and excited solutions. } 
\end{figure*}

Also note that the values of $M(0)$ generated in this covariant 
Yukawa theory show quite some hierarchy. In figure \ref{tressoluciones} 
one can see that they are well-spaced. 
Therefore this field-theory mechanism might conceivably be at work in
the fermion flavor problem.
However $M(0)$ seems to be larger for the nodeless solution than for 
the excited ones (this puzzling property also appears in the equal-time 
approach of \cite{Bicudo:2002eu}, and the extent of its model 
dependence needs to be further investigated). To confirm this point, we 
choose a different set of parameters and implementation of the 
computer code. 
With $m_\phi=100$, a cutoff $\Lambda=10^4$ implemented as a Gaussian 
fall-off of the integration measure (as opposed to terminating the 
grid), and $g=50$, with a small bare fermion mass $m_\psi=5\cdot 
10^{-4}$, one obtains the three solutions depicted in figure 
\ref{newsolplot}.
This graph confirms that the nodeless solution is higher in mass, and
that it is possible to generate sizeable mass hierarchies between the
solutions.  We use this run to also illustrate the dependence of $M_2$
with the cutoff, displayed in figure \ref{M2runswithg}.
As can be seen, the excited solution does not decrease monotonically as 
$g\to 0$, it rather seems to become a singular solution at the critical 
coupling. This technicality is however probably irrelevant for physical 
applications.

In hadron physics the strength of the interaction being fixed by 
extensive phenomenology, there would be no freedom to vary $g$.
 In the physically relevant cases studied in
\cite{Bicudo:2002eu}, excited solutions exist at physical values of the 
coupling.
In \cite{Nefediev:2004by} Ribeiro and Nefediev have further extended 
their original results to include bound states (mesons) constructed with 
the replicated quasiparticles. In particular there are Goldstone bosons 
over the replicated vacuum, that are of course not massless pions but 
would appear as excited pion states (separated from the ground state 
pion by the mass gap between the two vacua). Whether any conventional 
mesons accept a more convenient description as replicated mesons instead 
of excited mesons over the standard vacuum remains an open question.

Working in Euclidean space as we do provides fastly convergent 
integrals. 
This technicality  may be avoided by employing  a Lehmann 
representation \cite{Sauli:2002tk} that allows a direct solution in 
Minkowski space with similar results.

\section{Experimental signature}

In a free field theory, a field quantum of mass $m$  satisfies the 
equation
\be
E=\sqrt{{\bf p}^2+m^2}
\ee
that causes a pole in the free propagator
\be
S= \frac{i}{p^2-m^2} \ .
\ee

In an interacting field theory this pole will be shifted due to 
renormalization from its bare to its physical position. The dependence 
on the renormalization scale can be 
traded by a dependence on the particle Euclidean 4-momentum so that
\be
S= \frac{iZ(p^2)}{p^2-M(p^2)^2} 
\ee
yielding the transcendental equation for the pole position
\be
E=\sqrt{{\bf{p}}^2+M(E^2-{\bf{p}}^2)^2} \ .
\ee

\begin{figure*}
\psfig{figure=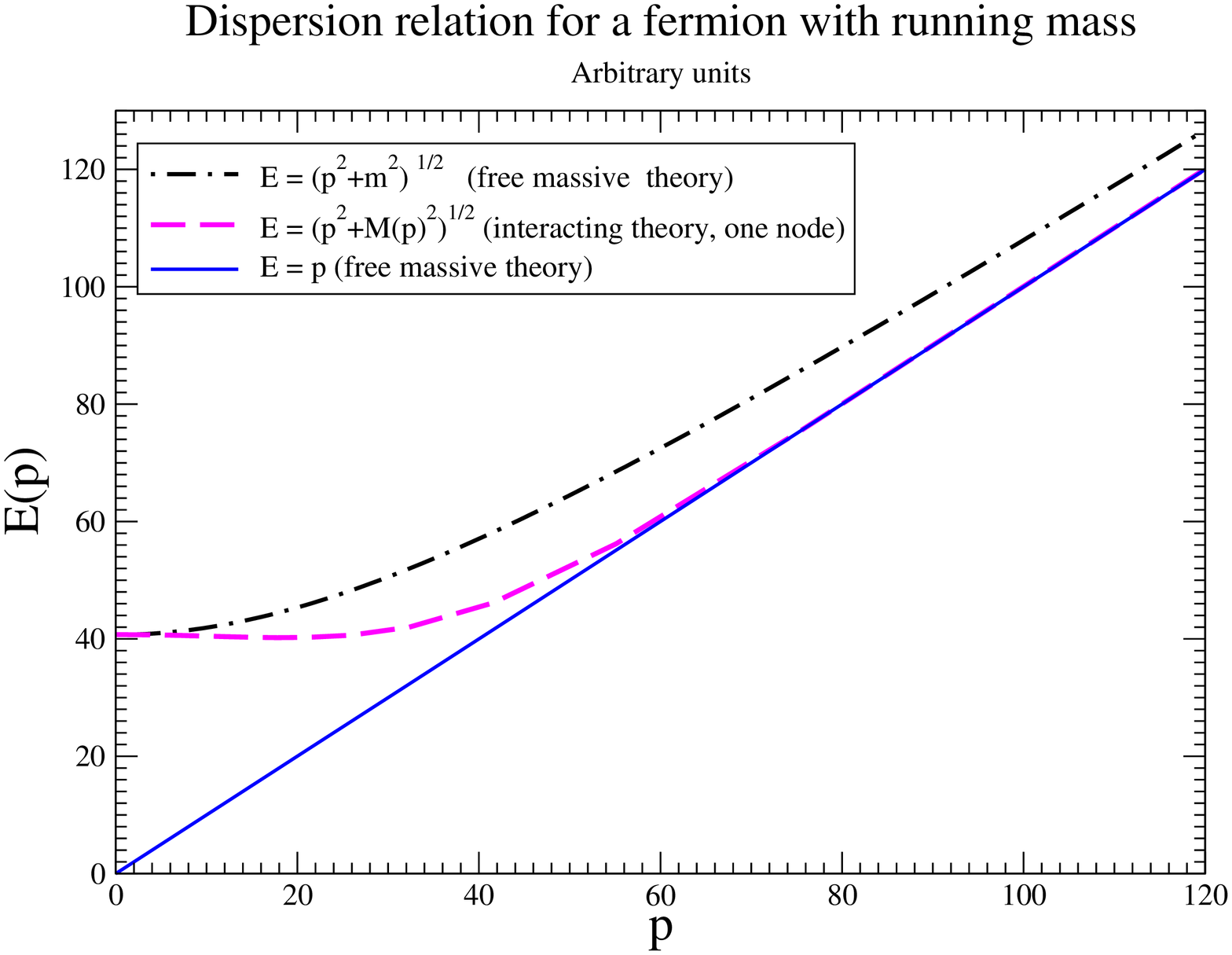,height=7in,angle=-90}
\caption{\label{dispersion2}
The virtual dispersion relation at fixed renormalization scale for the 
solution with two nodes (we find similar result for one node). As the 
running mass actually has a zero, the dispersive curve touches the massless 
limit that would otherwise be reached only asymptotically.
}
\end{figure*}

Another way of visualizing the function $M(p^2)$ of the interacting 
theory is to trade $p^2$ by the space-like part ${\bf p}^2$ as argument 
of $M$, generating therefore a non-trivial dispersion relation 
for virtual particles,
and the zeroes of the mass function $M$ appear as points where
$E=\ar {\bf p}\ar$ that would otherwise only be reached asymptotically 
at large $\bf p$ 
(see figure \ref{dispersion2}).
Tests for a non-trivial (real particle) dispersion relation have been 
proposed and some 
carried out in the search for violations of Lorentz invariance
\cite{Mattingly:2005re}. However these are very indirect and usually 
performed at low energy, so an accelerator-based test of the 
(virtual particle) dispersion 
relation by separately measuring $E$, $\bf{p}$ and the off-shellness 
is preferable.
In our case, real particles still provide representations of the 
Poincar\'e group and have constant, physical mass. However virtual 
particles off their mass shell will display the running mass function 
and this can be captured by analyzing the amplitudes of physical 
processes in perturbation theory.

\begin{figure*}
\psfig{figure=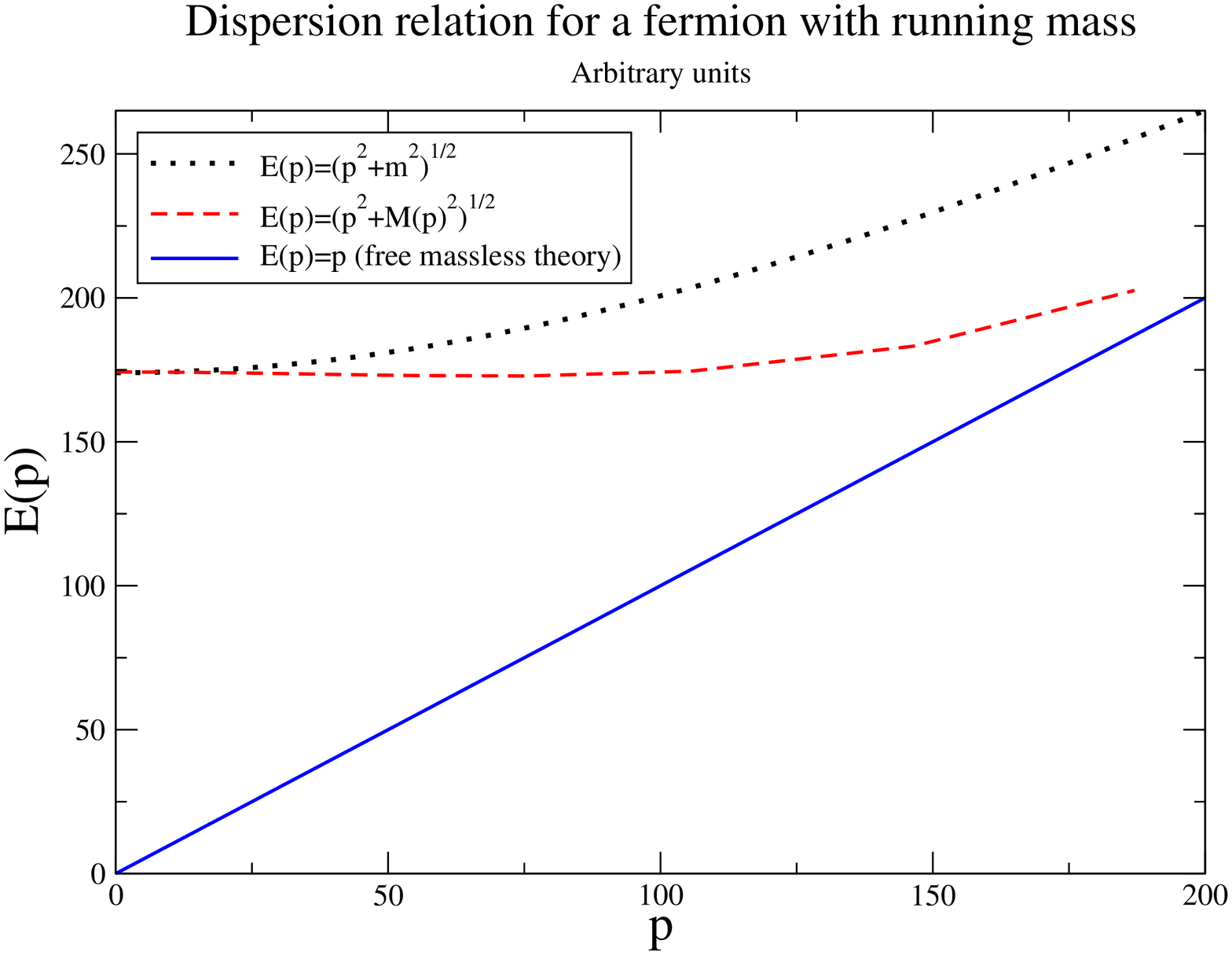,height=7in,angle=-90}
\caption{\label{dispersion1}
The virtual particle dispersion relation at fixed cutoff 
scale for the nodeless $M(p^2)$ solution. Note that the interacting 
theory falls 
below the usual Yukawa dispersion relation because the mass function 
drops with $p$. 
This is not very different from standard perturbative running with $p$
except the effect is much larger in the case of dynamical chiral 
symmetry breaking.}
\end{figure*}

The corresponding dispersion relations are plotted in figures
\ref{dispersion1} and \ref{dispersion2}.

Lacking a real theory there are two possibilities we cannot discern at 
present. Either all three families of fermions arise as solutions 
spontaneously breaking chiral symmetry, in which case two of them should 
already have a zero in their mass function, or 
the first generation reflects the trivial solution only slightly 
modifying the bare mass, and the second and third generation solutions 
with dynamically broken chiral symmetry, in which case only one of them 
would present a node.
In either case a zero is present in the mass function of one of the 
fermions. 

An interesting experiment is therefore a measurement of $\vec{p}$, $E$
and the off-shellness $\Delta^2 = p^2-M^2(p^2)$ of a $\tau$,  at various 
momenta $\vec{p}$. The $B$
factories have accumulated large $\tau$ samples~\cite{igonkina06} that
are under analysis and can be employed for this purpose.

The process one may investigate is depicted in
Fig.~\ref{fig:tauprod_offshell} in the appendix below. An 
electron-positron pair collides and
annihilates into a ${\tau^-}^* \tau^+$ pair, where the ${\tau^-}^*$ is
off its mass shell. Its four-momentum can be infered from the
center-of-momentum energy of the incoming $e^-e^+$ pair and by
reconstructing the energy and momentum of the on-shell $\tau^+$ from
its decay products, as detailed shortly. Finally, one needs the 
off-shellness, but this can be obtained from the number of counts with 
given $E$, $\ar{\bf p}\ar$.
This is obvious since the cross section for a specific process where the
off-shell $\tau^-$ decays into for example $l^- \bar{\nu}_l \nu_\tau$,
depends on the off-shellness $\Delta^2$ of the $\tau^-$ (see also
App.~\ref{sec:tau_prod}).

\begin{figure*}
\psfig{figure=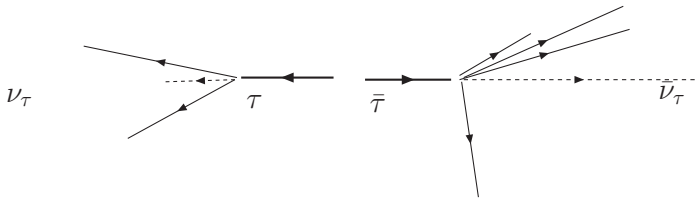,height=2in}
\caption{ In center of mass frame production where ${\bf p}_{\tau}+
{\bf p}_{\bar{\tau}}={\bf 0}$ one can fully reconstruct the decay of 
both $\tau$ leptons if only the neutrinos escape undetected. The 
analysis steps are described in the text.
} \label{taucm} 
\end{figure*}

Let us now detail the possible analysis:
\begin{enumerate}
\item In the center of mass frame of an $e^-e^+$ collision one can 
identify two back-to-back hadron jets, tagging the flavor by demanding 
that one of them kinematically reconstructs an on-shell $\tau$. This is 
rendered difficult by the undetected neutrino, that we assume to be the 
only unreconstructed track on the left side of the event. 
\item Complete reconstruction of the energy and momentum on the left 
side of the event (see Fig.~\ref{taucm}) is possible with a vertex 
detector (giving the direction of motion of the on-shell $\tau$). The 
total energy is taken from collider calibration, and matched to the 
energy of the visible tracks. Balancing energy provides the energy of 
the missing neutrino, and automatically the momentum of the left side, 
that can be tested against the hypothesis of a physical $\tau$ being 
produced.
\item Next one examines the right side of the reaction, where the 
secondary and primary vertex seem to coincide (as appropriate for a 
virtual, off-shell $\tau$ decaying rapidly). The total three-momentum is 
known a priori. One balances momentum to obtain the momentum taken by 
the undetected $\nu_\tau$, and obtains the total energy. Now, it is not 
necessarily true that $E^2-{\bf p}^2=m_\tau^2$.
\item The number of counts with given $E$, ${\bf p}$ on the right side 
is normalized to a cross section in terms of the 
off-shellness $\sigma(\Delta^2)$ and used 
to obtain $M$ from the value of the intermediate propagator in 
perturbation theory. 
\end{enumerate}


An alternative possibility is to carry out the measurement on the 
expected O(20-30) $\tau$ events at the OPERA experiment \cite{opera} 
in Gran Sasso National Laboratory. Given the momentum of the neutrinos 
from the CNGS beam one can attempt full kinematical reconstruction of 
$(E,{\bf p})$. However here the difficulty resides in the off-shellness
since a small value will be forced for the identification of the $\tau$.

The experimental test for the muon and the electron is even simpler, 
for example through Compton and reverse Compton scattering where the 
fermion is off-shell in the intermediate state. 
For quarks one needs to take into account that they always appear in bound 
states, and therefore the mass function is always under an integral sign 
and reconstruction from experimental data difficult. However a 
particularly simple case is 
$$
e^- e^+ \to \gamma^* \to s\bar{s}
$$
where the strangeness be completely tagged by counting all hyperons in 
both jets.

In principle the 
same analysis carried out here could also be undertaken for Majorana 
particles but this is out of our scope in this preliminary work.

\section{Conclusions}

We have called attention to an interesting feature of field theory, 
namely the possibility of having several solutions of the one-particle 
Dyson-Schwinger equations for a broad class of theories. The excited 
solutions appear upon increasing the coupling constant sufficiently, 
breaking chiral symmetry. 

The spectrum of states on top of each of the vacua is called ``replica'' 
or  ``recurrence'' and provides a mechanism that might be interesting 
for the theory of flavor. The number of recurrences arises from no 
special symmetry, since non-linear equations can perfectly have a finite 
number of solutions (again depending on the value of the coupling, this 
can be chosen to be three). 

These solutions have been shown in the past in a Hamiltonian framework 
\cite{Bicudo:2002eu} and we have minimally extended the results of these 
authors to show that the covariant formulation allows similar 
phenomena.
The excited solutions have mass functions that present zeroes and this 
reflects in the virtual dispersion relation (feature of an interacting 
field 
theory that should not be confused with breaking of Lorentz invariance).

Whether these solutions will be relevant to the fermion family problem 
can be tested at $B$ factories employing their large $\tau$ samples 
directly produced.

At this point it is a meaningless exercise to attempt to obtain the 
fermion masses in this scheme. Three data points $M_i(0)$ should be 
used to fit three parameters, the current fermion mass (that we have 
here set to zero), the boson mass, and $g$ at a given cutoff, so
the current predictive power is null.
However we find this field theory mechanism still worth attention given 
the ease with which fermion coupling universality can be incorporated 
by having three recurrences of the same field. 

Testing the off-shell dispersion relation should allow to immediately 
discard the scenario so we believe we have presented a valid physical 
hypothesis.
Of course, the success of radiative corrections in the standard model 
make one wish for a measurement at the highest possible energy, maybe at 
a future linear collider.

{\it This work has been performed in the framework of the research
projects FPA 2004-02602, 2005-02327, PR27/05-13955-BSCH (Spain) and is
part of the Masters thesis of Mr. P\'aramo Mart\'{\i}n presented to
the faculty of U. Complutense). TVC is a postdoctoral fellow for the
Fund for Scientific Research - Flanders and acknowledges the support
of the ``Programa de Investigadores Extranjeros en la UCM - Grupo
Santander''.}

\appendix
\section{$\tau$-pair production in $e^+e^-$ collisions}\label{sec:tau_prod}
\subsection{On-shell cross section}\label{sec:onshell}

\begin{figure*}
\begin{center}
\resizebox*{5cm}{!}{\includegraphics[width=5cm]{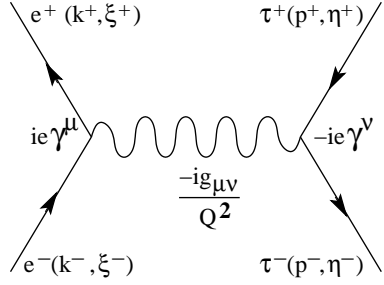}}
\caption{On-shell $\tau$-pair production in $e^+e^-$ collisions. The
  particles' momenta and spins are indicated, as well as the
  conventions for the photon propagator and the vertices.}
\label{fig:tauprod_onshell}
\end{center}
\end{figure*}
We look at the $e^+e^- \to \tau^+ \tau^-$ process depicted in
Fig.~\ref{fig:tauprod_onshell}, where all ingoing and outgoing
particles are on their mass shell. Using the conventions of
Ref.~\cite{halzenmartin84}, the amplitude for the depicted process is
\begin{multline}
-i \mM = \ubar^{\tau^-}_{\eta^-} (p^-) \ \left( -i e \gamma^\nu
\right) \ v^{\tau^+}_{\eta^+} (p^+) \; \left(
\frac{-ig_{\mu\nu}}{Q^2} \right) \\ \times \; \vbar^{e^+}_{\xi^+} (k^+) \
\left( i e \gamma^\mu \right) \ u^{e^-}_{\xi^-} (k^-) \; . \;
\label{eq:amp_onshell}
\end{multline}

For the unpolarized cross section, one needs the squared amplitude
averaged over initial and summed over final spins
\begin{multline}
\overset{\rule{6mm}{.2mm}}{\sum} |\mM|^2 = \frac{1}{4} \frac{e^4}{s^2}
\sum_{\textrm{all spins}} \ubar^{\tau^-}_{\eta^-} (p^-)
\, {\gamma_\mu}_{\eta^-\eta^+} \, v^{\tau^+}_{\eta^+} (p^+)
\\ \times \, \vbar^{e^+}_{\xi^+} (k^+) \, {\gamma^\mu}_{\xi^+\xi^-} \,
u^{e^-}_{\xi^-} (k^-) \, \ubar^{e^-}_{\xi'^-} (k^-) \, {\gamma^\nu}_{\xi'^-\xi'^+} \,
v^{e^+}_{\xi'^+} (k^+) \\ \times \, \vbar^{\tau^+}_{\eta'^+} (p^+) \,
{\gamma_\nu}_{\eta'^+\eta'^-} \, u^{\tau^-}_{\eta'^-} (p^-) \; , \;
\label{eq:amp_sum}
\end{multline}
where the spinor indices are explicitly shown. For $e^+e^-$
annihilation, one has $Q^2=s$. One can now rearrange the factors in
Eq.~(\ref{eq:amp_sum}). Making use of the relations
\begin{subequations}
\begin{eqnarray}
u^{l}_{\xi} (p) \ubar^{l}_{\xi'} (p) & = & \left( \psl + m_l
\right)_{\xi \xi'} \; , \; \label{eq:clos_u_spinor} \\
v^{l}_{\xi} (p) \vbar^{l}_{\xi'} (p) & = & \left( \psl - m_l
\right)_{\xi \xi'} \; , \; \label{eq:clos_v_spinor} 
\end{eqnarray}
\label{eq:clos_spinor}
\end{subequations}
one can write Eq.~(\ref{eq:amp_sum}) as a product of traces
\begin{multline}
\overset{\rule{6mm}{.2mm}}{\sum} |\mM|^2 \; = \; \frac{1}{4} \;
\frac{e^4}{s^2} \; \textrm{Tr} \left[ \left( \ksl^- + m_e \right)
  \gamma^\nu \left( \ksl^+ - m_e \right) \gamma^\mu \right] \\
\times \textrm{Tr} \left[ \left( \psl^- + m_\tau \right) \gamma_\mu
  \left( \psl^+ - m_\tau \right) \gamma_\nu \right] \; . \;
\label{eq:amp_tr}
\end{multline}
Neglecting the electron mass, this results in
\begin{multline}
\overset{\rule{6mm}{.2mm}}{\sum} |\mM|^2 \; = \; \frac{8e^4}{s^2} \;
\bigl( k^- \cdot p^- \, k^+ \cdot p^+ \; + \; k^- \cdot p^+ \, k^+
\cdot p^- \phantom{\bigr)} \\ \phantom{\bigl(} \; + \; m^2_\tau \, k^-
\cdot k^+ \bigr) \; . \;
\label{eq:amp_4vecprod}
\end{multline}
This can be written in terms of Mandelstam variables
\begin{subequations}
\begin{eqnarray}
s & = & (k^- - k^+)^2 = (p^+ - p^-)^2 = Q^2 \; , \,
\label{eq:mand_s}\\
t & = & (k^- - p^-)^2 = (p^+ - k^+)^2 \; , \, \label{eq:mand_t}\\
u & = & (k^- - p^+)^2 = (p^- - k^+)^2 \; , \, \label{eq:mand_u}
\end{eqnarray}
\label{eq:mand_var}
\end{subequations}
as
\begin{equation}
\overset{\rule{6mm}{.2mm}}{\sum} |\mM|^2 \; = \; \frac{2e^4}{s^2} \;
\left( t^2 \; + \; u^2 \; + \; 4 \, m^2_\tau \, s \; - \; 2 \,
m^4_\tau \right) \; , \;
\label{eq:amp_mand}
\end{equation}
where at high energies, the mass terms become negligible. In the
center-of-momentum (com) frame, the amplitude is
\begin{multline}
\overset{\rule{6mm}{.2mm}}{\sum} |\mM|^2 \; = \; \frac{e^4}{s^2} \;
\bigl[ s^2 \, \left( 1 \; + \; \cos^2(\theta_\tau) \right)
  \phantom{\bigr]} \\ \phantom{\bigl[} \; + \; 4
\, m^2_\tau \, s \, \left( 1 \; - \cos^2(\theta_\tau) \right) \bigr]
\; , \; \label{eq:amp_com}
\end{multline}
with $\theta_\tau$ the com-angle between incoming electron $e^-$ and
outgoing $\tau^-$.

The differential cross section in the com-frame is then given by
\begin{multline}
\phantom{\Bigl|} {\frac{\ud \sigma}{\ud \Omega}} \Bigr|_{\textrm{com}}
\; = \; \frac{\alpha^2}{4s^2} \, \sqrt{1 - \frac{4m^2_\tau}{s}} \\
\times \, \left[ \, s \left( 1 + \cos^2(\theta_\tau) \right) + 4 m^2_\tau
\sin^2(\theta_\tau) \right] \; , \; \label{eq:diffcs_com}
\end{multline}
with $\alpha = e^2 / 4\pi$. The total cross section is
\begin{equation}
\sigma^{\textrm{com}}_{\textrm{tot.}} \; = \; \frac{4 \pi \alpha^2}{3
  s} \, \sqrt{1 - \frac{4m^2_\tau}{s}} \, \left( 1 + \frac{2 
m^2_\tau}{s}
  \right) \; . \; \label{eq:cs_tot_com}
\end{equation}

\subsection{Half off-shell cross section}\label{sec:ofshell}
\begin{figure*}
\begin{center}
\resizebox*{10cm}{!}{\includegraphics{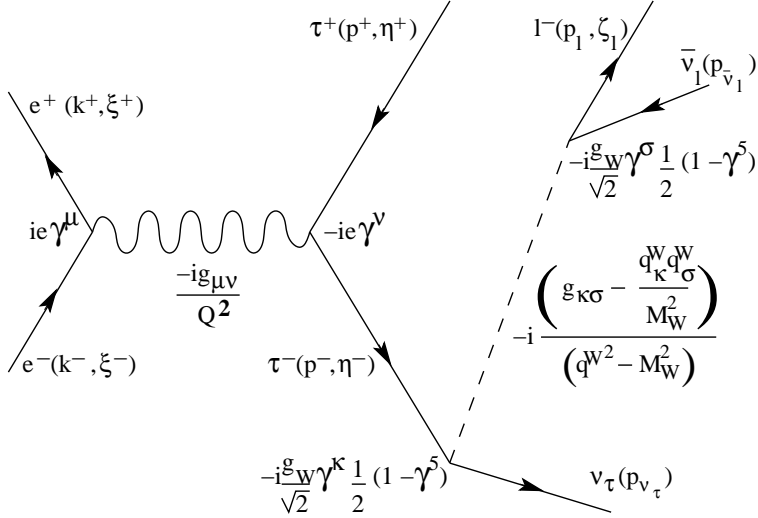}}
\caption{Half off-shell $\tau$-pair production in $e^+e^-$
  collisions. The particles' momenta and spins are indicated, as well
  as the conventions for the boson propagators and the vertices.}
\label{fig:tauprod_offshell}
\end{center}
\end{figure*}
In this part of the Appendix, we look at the $e^-e^+ \to \tau^-
\tau^+$ process, where one of the $\tau$-particles is off its mass
shell and decays into a lighter lepton $l$ (electron or muon). We
discuss the case for an off-shell $\tau^-$. The case for an off-shell
$\tau^+$ is clearly completely analogous. The process is displayed in
Fig.~\ref{fig:tauprod_offshell}.

The off-shellness of one of the $\tau$'s changes the kinematics. The
total energy in the reaction is unevenly divided: $E_{\tau^+} \neq
E_{\tau^-} \neq W_{\textrm{com}}/2$, where $W_{\textrm{com}} =
\sqrt{s}$ is the total energy in the com-frame. The amplitude for the
process depicted in Fig.~\ref{fig:tauprod_offshell} is
\begin{multline}
-i \mM \, = \, \ubar^{l^-}_{\zeta_l} (p_l) \left( -i \,
\frac{g_W}{\sqrt{2}} \, \gamma^\sigma \frac{1}{2} \left( 1 - \gamma^5
\right) \right) v^{\bar{\nu}_l} (p_{\bar{\nu}_l}) \\ \times \, \left( -i \right)
\frac{\left( g_{\kappa \sigma} - \frac{q^W_\kappa q^W_\sigma} {M^2_W}
  \right)} {\left( {q^W}^2 - M^2_W \right)} \, \ubar^{\nu_\tau}
(p_{\nu_\tau}) \left( -i \, \frac{g_W}{\sqrt{2}} \, \gamma^\sigma
\frac{1}{2} \left( 1 - \gamma^5
\right) \right) \\ \times \, \frac{i}{\left( \psl^- - M_{\tau^-} \right)} \,
\left( -i e \gamma^\nu \right) \ v^{\tau^+}_{\eta^+} (p^+) \, \left(
\frac{-ig_{\mu\nu}}{q^2} \right) \\ \times \,
\vbar^{e^+}_{\xi^+} (k^+) \ \left( i e \gamma^\mu \right) \
u^{e^-}_{\xi^-} (k^-) \; . \;
\label{eq:amp_offshell}
\end{multline}

The propagator of the off-shell $\tau^-$ will give rise to a
dependence of the amplitude on the \textit{off-shellness} $\Delta^2 =
E^2_{\tau^-} - \vec{p}^2_\tau - M^2_{\tau^-}$.

In the limit of infinite mass of the $W$-boson and vanishing electron
mass, the unpolarized squared amplitude for the process depicted in
Fig.~\ref{fig:tauprod_offshell} is~\cite{form}
\begin{multline}
\overset{\rule{6mm}{.2mm}}{\sum} |\mM|^2 =
\frac{2^{12} \ \pi^4 \ \alpha^2_{\textrm{QED}} \
  \alpha^2_{\textrm{weak}}} {M^4_W \ \Delta^4 \ q^4} \; p_{\nu_\tau}
\cdot p_l \\
\times \Bigl[ \ p_{\bar{\nu}_l} \cdot p^-
\, \bigl( \ m_\tau \ M_{\tau^-} \ k^+ \cdot k^- \; + \; k^+ \cdot p^+
\ k^- \cdot p^- \; + \; k^+ \cdot p^- \ k^- \cdot p^+ \bigr)
\phantom{\Bigr]} \\
\phantom{\Bigl[} - \; \frac{\Delta^2}{2} \; \bigl( \ k^+
\cdot p_{\bar{\nu}_l} \ k^- \cdot p^+ \; + \; k^+ \cdot p^+ \ k^- \cdot
p_{\bar{\nu}_l} \ \bigr) \Bigr] \; . \;
\label{eq:amp_sq_off}
\end{multline}

The (differential) cross section for this process is proportional to
the above squared amplitude, integrated over the threemomenta of the
outgoing $\tau$-neutrino and lepton-antineutrino. This is a standard 
computation that can be incorporated in the computerr code if needed, 
yet it is clear that the resulting cross section will depend on
the off-shellness $\Delta^2$ and the virtual $\tau$ mass-energy
dispersion relation. In particular, at large off-shellness, the cross 
section behaves as $1/\Delta^2$ and the constant multiplying this 
parametric behavior can be fit to data.

\end{document}